\documentclass[
reprint,  
superscriptaddress,
 amsmath,amssymb,
 aps,
 prl,
 bibnotes
]{revtex4-2}

\usepackage{graphicx}
\usepackage{dcolumn}
\usepackage{bm}
\usepackage{hyperref}
\usepackage{multirow}
\usepackage{color}
\usepackage{soul}

\begin{document}
 \textit{Preprint published in Physical Review Letters as DOI:} \url{https://doi.org/10.1103/d7dr-h36j}
\date{\today}

\title{\textbf{Constraining the Synthesis of the Lightest \textit{p} Nucleus $^{74}$Se} 
}%

\author{A. Tsantiri}\email{artemis.tsantiri@uregina.ca}
\altaffiliation{Present address: Department of Physics, University of Regina, Regina, SK S4S 0A2, Canada}
\affiliation{Facility for Rare Isotope Beams, Michigan State University, East Lansing, Michigan 48824, USA}
\affiliation{Department of Physics and Astronomy, Michigan State University, East Lansing, Michigan 48824, USA}
\affiliation{NuGrid Collaboration, \url{http://nugridstars.org}}

\author{A. Spyrou}\email{spyrou@frib.msu.edu}
\affiliation{Facility for Rare Isotope Beams, Michigan State University, East Lansing, Michigan 48824, USA}
\affiliation{Department of Physics and Astronomy, Michigan State University, East Lansing, Michigan 48824, USA}

\author{E. C. Good}
\altaffiliation{Present address: Pacific Northwest National Laboratory, Richland, Washington 99352, USA}
\affiliation{Facility for Rare Isotope Beams, Michigan State University, East Lansing, Michigan 48824, USA}

\author{K. Bosmpotinis}
\affiliation{Facility for Rare Isotope Beams, Michigan State University, East Lansing, Michigan 48824, USA}
\affiliation{Department of Physics and Astronomy, Michigan State University, East Lansing, Michigan 48824, USA}

\author{P. Giuliani}
\affiliation{Facility for Rare Isotope Beams, Michigan State University, East Lansing, Michigan 48824, USA} 

\author{H. Arora}
\affiliation{Department of Physics, Central Michigan University, Mount Pleasant, Michigan 48859, USA}

\author{G. Balk}
\affiliation{Physics Department, Hope College, Holland, Michigan 49423, USA}

\author{L. Balliet}
\affiliation{Facility for Rare Isotope Beams, Michigan State University, East Lansing, Michigan 48824, USA}
\affiliation{Department of Physics and Astronomy, Michigan State University, East Lansing, Michigan 48824, USA}

\author{H. C. Berg}
\affiliation{Facility for Rare Isotope Beams, Michigan State University, East Lansing, Michigan 48824, USA}
\affiliation{Department of Physics and Astronomy, Michigan State University, East Lansing, Michigan 48824, USA}

\author{J. M. Berkman}
\affiliation{Facility for Rare Isotope Beams, Michigan State University, East Lansing, Michigan 48824, USA}
\affiliation{Department of Chemistry, Michigan State University, East Lansing, Michigan 48824, USA}

\author{C. Dembski}
\affiliation{Department of Physics, University of Notre Dame, Notre Dame, Indiana 46556, USA}

\author{P. DeYoung}
\affiliation{Physics Department, Hope College, Holland, Michigan 49423, USA}

\author{P. A. Denissenkov}
\affiliation{Department of Physics and Astronomy, University of Victoria, Victoria, British Columbia, V8W~2Y2, Canada}
\affiliation{CaNPAN, \url{https://canpan.ca}}
\affiliation{NuGrid Collaboration, \url{http://nugridstars.org}}

\author{N. Dimitrakopoulos}
\affiliation{Department of Physics, Central Michigan University, Mount Pleasant, Michigan 48859, USA}

\author{A. Doetsch}
\affiliation{Facility for Rare Isotope Beams, Michigan State University, East Lansing, Michigan 48824, USA}
\affiliation{Department of Physics and Astronomy, Michigan State University, East Lansing, Michigan 48824, USA}

\author{T. Gaballah}
\affiliation{Department of Physics and Astronomy, Mississippi State University, Mississippi State, Mississippi 39762, USA}
\author{R. Garg}
\affiliation{Facility for Rare Isotope Beams, Michigan State University, East Lansing, Michigan 48824, USA} 

\author{A. Henriques}
\affiliation{Facility for Rare Isotope Beams, Michigan State University, East Lansing, Michigan 48824, USA}

\author{R. Jain}
\altaffiliation{Present address: Lawrence Livermore National Laboratory, 7000 East Avenue, Livermore, California 94550-9234, USA}
\affiliation{Facility for Rare Isotope Beams, Michigan State University, East Lansing, Michigan 48824, USA}
\affiliation{Department of Physics and Astronomy, Michigan State University, East Lansing, Michigan 48824, USA}

\author{S. N. Liddick}
\affiliation{Facility for Rare Isotope Beams, Michigan State University, East Lansing, Michigan 48824, USA}
\affiliation{Department of Chemistry, Michigan State University, East Lansing, Michigan 48824, USA}

\author{S. Lyons}
\affiliation{Pacific Northwest National Laboratory, Richland, Washington 99352, USA}

\author{R. S. Lubna}
\affiliation{Facility for Rare Isotope Beams, Michigan State University, East Lansing, Michigan 48824, USA}

\author{B. Monteagudo Godoy}
\affiliation{Physics Department, Hope College, Holland, Michigan 49423, USA}

\author{F. Montes}
\affiliation{Facility for Rare Isotope Beams, Michigan State University, East Lansing, Michigan 48824, USA}

\author{S. Nash}
\affiliation{Facility for Rare Isotope Beams, Michigan State University, East Lansing, Michigan 48824, USA}

\author{G. U. Ogudoro}
\affiliation{Physics Department, Hope College, Holland, Michigan 49423, USA}

\author{J. Owens-Fryar}
\affiliation{Facility for Rare Isotope Beams, Michigan State University, East Lansing, Michigan 48824, USA}
\affiliation{Department of Physics and Astronomy, Michigan State University, East Lansing, Michigan 48824, USA}

\author{A. Palmisano-Kyle}
\affiliation{Department of Physics and Astronomy, University of Tennessee, Knoxville, Tennessee 37996, USA}

\author{J. Pereira}
\affiliation{Facility for Rare Isotope Beams, Michigan State University, East Lansing, Michigan 48824, USA}

\author{A. Psaltis}
\affiliation{Department of Physics, Duke University, Durham, NC 27710, USA}
\affiliation{Triangle Universities Nuclear Laboratory, Duke University, Durham, NC 27710, USA}
\affiliation{NuGrid Collaboration, \url{http://nugridstars.org}}

\author{A. L. Richard}
\affiliation{Department of Physics and Astronomy, Ohio University, Athens, Ohio 45701, USA}
\affiliation{Facility for Rare Isotope Beams, Michigan State University, East Lansing, Michigan 48824, USA}

\author{L. Roberti}
\affiliation{Istituto Nazionale di Fisica Nucleare-Laboratori Nazionali del Sud, Via Santa Sofia 62, Catania, I-95123, Italy}
\affiliation{Konkoly Observatory, Research Centre for Astronomy and Earth Sciences, Konkoly-Thege Mikl\'os \'ut 15-17, 1121 Budapest, Hungary}
\affiliation{CSFK HUN-REN, MTA Centre of Excellence, Konkoly Thege Mikl\'os \'ut 15-17, Budapest, H-1121, Hungary}
\affiliation{Istituto Nazionale di Astrofisica–Osservatorio Astronomico di Roma, Via Frascati 33, Monte Porzio Catone, I-00040, Italy}
\affiliation{NuGrid Collaboration, \url{http://nugridstars.org}}

\author{E. K. Ronning}
\affiliation{Facility for Rare Isotope Beams, Michigan State University, East Lansing, Michigan 48824, USA}
\affiliation{Department of Chemistry, Michigan State University, East Lansing, Michigan 48824, USA}

\author{H. Schatz}
\affiliation{Facility for Rare Isotope Beams, Michigan State University, East Lansing, Michigan 48824, USA}
\affiliation{Department of Physics and Astronomy, Michigan State University, East Lansing, Michigan 48824, USA}

\author{A. Sebastian}
\affiliation{Facility for Rare Isotope Beams, Michigan State University, East Lansing, Michigan 48824, USA}
\affiliation{Department of Physics and Astronomy, Michigan State University, East Lansing, Michigan 48824, USA}

\author{M. Smith}
\affiliation{Facility for Rare Isotope Beams, Michigan State University, East Lansing, Michigan 48824, USA}
\affiliation{Department of Physics and Astronomy, Michigan State University, East Lansing, Michigan 48824, USA}

\author{M. K. Smith}
\affiliation{Facility for Rare Isotope Beams, Michigan State University, East Lansing, Michigan 48824, USA}

\author{C. S. Sumithrarachchi}
\affiliation{Facility for Rare Isotope Beams, Michigan State University, East Lansing, Michigan 48824, USA}

\author{C. Tinson}
\affiliation{Facility for Rare Isotope Beams, Michigan State University, East Lansing, Michigan 48824, USA}
\affiliation{Department of Physics and Astronomy, Michigan State University, East Lansing, Michigan 48824, USA}

\author{P. Tsintari}
\altaffiliation{Present address: Facility for Rare Isotope Beams, Michigan State University, East Lansing, Michigan 48824, USA}
\affiliation{Department of Physics, Central Michigan University, Mount Pleasant, Michigan 48859, USA}

\author{N. Tubaro}
\altaffiliation{Present address: Department of Physics, University of Notre Dame, Notre Dame, Indiana 46556, USA}
\affiliation{Facility for Rare Isotope Beams, Michigan State University, East Lansing, Michigan 48824, USA}
\affiliation{Department of Physics and Astronomy, Michigan State University, East Lansing, Michigan 48824, USA}

\author{S. Uthayakumaar}
\affiliation{Facility for Rare Isotope Beams, Michigan State University, East Lansing, Michigan 48824, USA}

\author{A.C.C. Villari}
\affiliation{Facility for Rare Isotope Beams, Michigan State University, East Lansing, Michigan 48824, USA}

\author{E. Weissling}
\affiliation{Physics Department, Hope College, Holland, Michigan 49423, USA}

\author{R. G. T. Zegers}
\affiliation{Facility for Rare Isotope Beams, Michigan State University, East Lansing, Michigan 48824, USA}
\affiliation{Department of Physics and Astronomy, Michigan State University, East Lansing, Michigan 48824, USA}

\begin{abstract}
We provide the first experimental cross section of the $^{73}\text{As}(p,\gamma)^{74}\text{Se}$ reaction to constrain one of the main destruction mechanisms of the \textit{p} nucleus $^{74}\text{Se}$ in explosive stellar environments. The measurement was done using a radioactive $^{73}\text{As}$ beam at effective center-of-mass energies of 2.9 and 2.3  MeV/nucleon. Along with the total cross-section measurement, statistical properties of the $^{74}\text{Se}$ compound nucleus were extracted, constraining the reaction cross section in the upper Gamow window of the $\gamma$ process. The impact of the experimentally constrained reaction rate on $^{74}\text{Se}$ production in Type II supernovae was investigated through Monte Carlo one-zone network simulations. The results indicate that the overproduction of $^{74}$Se by Type II supernova models cannot be resolved by nuclear physics alone and point toward the need for a more detailed understanding of the astrophysical conditions of relevance for the $\gamma$ process.
\end{abstract}

\maketitle

Beyond iron, there is a group of rare, stable, neutron-deficient isotopes between $^{74}\text{Se}$ and $^{196}\text{Hg}$ known as the \textit{p} nuclei. Even though they constitute only a small fraction of the heavy nuclear species, understanding their production mechanism has been a fundamental question in nuclear astrophysics since the field's inception \cite{b2fh} and remains a focus of research today \cite{battino2020,Choplin_2022,roberti2023}. The most widely accepted scenario for the formation of \textit{p} nuclei is the $\gamma$ process, which involves the disintegration of preexisting neutron-rich heavy seed material through a series of photon-induced reactions \cite{Arnould1976, Rauscher_2013, pignatari_2016_review}. This process is thought to occur in the ONe layers of Type II core collapse supernovae (CCSN) \cite{Woosley_1978,Prantzos_1990} and in thermonuclear Type Ia supernovae \cite{Travaglio_2011, Kusakabe_2011, battino2020}. In such environments, high temperatures of $T \sim 2 - 3.5$ GK guide the nuclear flux through consecutive $(\gamma, n)$ reactions toward unstable, proton-rich species, followed by a combination of $(\gamma, p)$ or $(\gamma, \alpha)$ reactions. After the explosion, the produced radioactive isotopes decay back to stability via $\beta^+$/electron capture decays, reaching the elusive \textit{p} nuclei. As p-nuclei abundances cannot be derived from stellar spectra, our knowledge of their isotopic distribution in the Solar System comes primarily from primitive meteorites, which preserve a near-solar composition for many heavy elements, serving as the main benchmark in testing $\gamma$-process models \cite{Rauscher_2013, pignatari_2016_review}.

Modeling the $\gamma$ process requires a complex network of thousands of nuclear reactions, most of which involve radioactive nuclei. Due to the scarcity of experimental cross sections, reaction rates are typically determined from Hauser-Feshbach statistical model calculations \cite{hauser_feshbach}. However, those predictions become increasingly uncertain for reactions involving nuclei further from stability \cite{Rauscher_2013, Arnould_2003}. While several indirect approaches can help constrain these rates, direct measurements remain the most reliable since they best resemble the stellar environment where the reactions occur. It is therefore essential to perform direct measurements to reduce these uncertainties and improve the accuracy of theoretical models. Despite extensive experimental efforts over the past decades to measure cross sections relevant to the $\gamma$ process on stable nuclei \cite{Gyürky_2007, Gyürky_2014, Mei_2015,Harissopulos_2016, Psaltis_2019, Simon_2019,Banu_2019, Heim_2020, Cheng_2021, review_gn_ELI-NP}, until today there has been only one measurement involving a radioactive beam within the Gamow window \cite{Lotay_2021} ($\sim1.3-4.0$ MeV in the center-of-mass system~\cite{Iliadis}) and two at energies higher than those relevant for the $\gamma$ process \cite{Glorius_2019, Dellmann_2025}. The combination of small cross sections and low radioactive beam intensities makes such measurements particularly challenging, and therefore, they remain an active pursuit at many radioactive beam facilities. In this Letter, we present the first measurement of the $^{73}\text{As}(p,\gamma)^{74}\text{Se}$ cross section within the $\gamma$-process Gamow window using a radioactive $^{73}\text{As}$ beam and investigate its impact on the production of the lightest \textit{p} nucleus, $^{74}\text{Se}$. 
  
While CCSN models \cite{rauscher_2002_ccsn_models, Pignatari_2016_setI, ritter_2018_setII,lawson_2022} show variations in the production of the \textit{p} nuclei, $^{74}\text{Se}$ is often found to be overproduced compared to solar abundances \cite{lodders}. To explore whether this overproduction arises from reaction-rate uncertainties, the production and destruction mechanisms of $^{74}\text{Se}$ have been a topic of study for more than a decade. Specifically, the $^{74}\text{Se}(p,\gamma)^{75}\text{Br}$ \cite{krivonosov_1977_Se_pg, guyrky_2003_se_pg, skakun_2010_Se_pg, foteinou_2018_se_pg}, $^{70}\text{Ge}(\alpha,\gamma)^{74}\text{Se}$ \cite{fulop_1996_70Ge_ag_74Se}, and $^{74}\text{Se}(n,\gamma)^{75}\text{Se}$ \cite{dillmann_2006_74se_ng_75Se} reactions have been measured directly. The last unmeasured reaction that significantly affects the final $^{74}\text{Se}$ abundance is the $^{74}\text{Se}(\gamma, p)^{73}\text{As}$, which has been identified as a key reaction to destroy $^{74}\text{Se}$ \cite{rapp06}.

As matter in the stellar plasma is thermally excited, it is often advantageous to experimentally measure the radiative capture reactions of positive $Q$ value, in this case the $^{73}\text{As}(p,\gamma)^{74}\text{Se}$ reaction, to reduce the stellar enhancement effect \cite{Rauscher_2009}. The rate of their photon-induced counterparts can then be inferred through the detailed balance theorem \cite{Iliadis}. In CCSN models \cite{rauscher_2002_ccsn_models, Pignatari_2016_setI, ritter_2018_setII}, $^{74}\text{Se}$ is found to be mostly produced near the upper end of the $\gamma$-process temperature range, around 3 GK. The corresponding Gamow window for the $^{73}\text{As}(p,\gamma)^{74}\text{Se}$ reaction is between 1.7 and 3.5 MeV \cite{Iliadis}. 

In this energy region, the theoretically predicted cross-section uncertainty arises primarily from uncertainties in the statistical properties of the $^{74}\text{Se}$ compound nucleus, namely, the nuclear level density (NLD) and $\gamma$-strength function ($\gamma$SF). This uncertainty can be quantified with open source statistical modeling codes such as \textsc{Talys}~\cite{talys}. For this particular reaction, the uncertainty calculated using default NLD and $\gamma$SF model parameters available within \textsc{Talys} varies by a factor of $\approx$ 6. However, it has been seen that, particularly for unstable isotopes, this approach may underestimate the theoretical uncertainty~\cite{dennis_shape, spyrou_ba}. In this Letter, we provide the first cross-section measurement of the $^{73}\text{As}(p,\gamma)^{74}\text{Se}$ reaction within the Gamow window, as well as constraints in the statistical properties of $^{74}\text{Se}$, significantly reducing the uncertainty of the final production of $^{74}\text{Se}$ in CCSN.
 
The experiment took place at the Facility for Rare Isotope Beams, Michigan State University. A radioactive $^{73}\text{As}$ beam was produced from a source sample evaporated in the batch-mode ion source \cite{candia_73As_beam, SUMITHRARACHCHI2023301}. The beam was charge bred to the 23+ charge state, accelerated to 3.1 and 3.7 MeV/nucleon in the ReA reaccelerator facility \cite{ReA}, and delivered onto a hydrogen gas-cell target in the experimental end station. The $\gamma$ rays produced by the reaction were detected using the Summing NaI(Tl) (SuN) detector \cite{SuN_paper2013} and were analyzed using the $\gamma$-summing technique \cite{tas}. A detailed description of the experimental setup and analysis technique can be found in Refs. \cite{tsantiri, palmisano} from the proof-of-principle experiment, and the main components are briefly summarized here.

The target was a 4-cm-long gas cell made of plastic positioned at the center of the SuN detector. It had a 2-$\mu$m-thick molybdenum (Mo) entrance window, a 5-$\mu$m-thick exit window, and was filled with hydrogen gas at a pressure of $\simeq$ 600 Torr. The interior and the front surfaces of the cell were lined with tantalum to minimize beam interaction with the plastic and to reduce background. Surrounding the target was the barrel-shaped SuN detector \cite{SuN_paper2013}. SuN is divided into eight optically isolated segments and can provide three main types of spectra: total absorption spectra (TAS), sum of segments (SoS), and multiplicity. TAS corresponds to the full energy deposited in the detector, SoS corresponds to the sum of the energy spectra recorded by each segment, and multiplicity indicates how many segments recorded energy in each event \cite{tas}.

Proton capture on the $^{73}\text{As}$ beam forms a $^{74}\text{Se}$ compound nucleus at an excited state $E_X = Q + E_{CM}$, where $Q=8.549(4)$ MeV \cite{AME2020} and $E_{CM}$ is the total kinetic energy in the center-of-mass system. Adding all $\gamma$ rays originating from a single cascade forms the so-called ``sum peak" in the TAS spectrum with energy equal to $E_X$. The integral of the sum peak corresponds to the number of reactions detected, which represents the ratio of the number of reactions taking place and the number of projectiles impinging on the target, namely, the experimental yield, $Y$ \cite{SuN_paper2013}. 

\begin{figure}[t]
    \includegraphics[width=\linewidth]{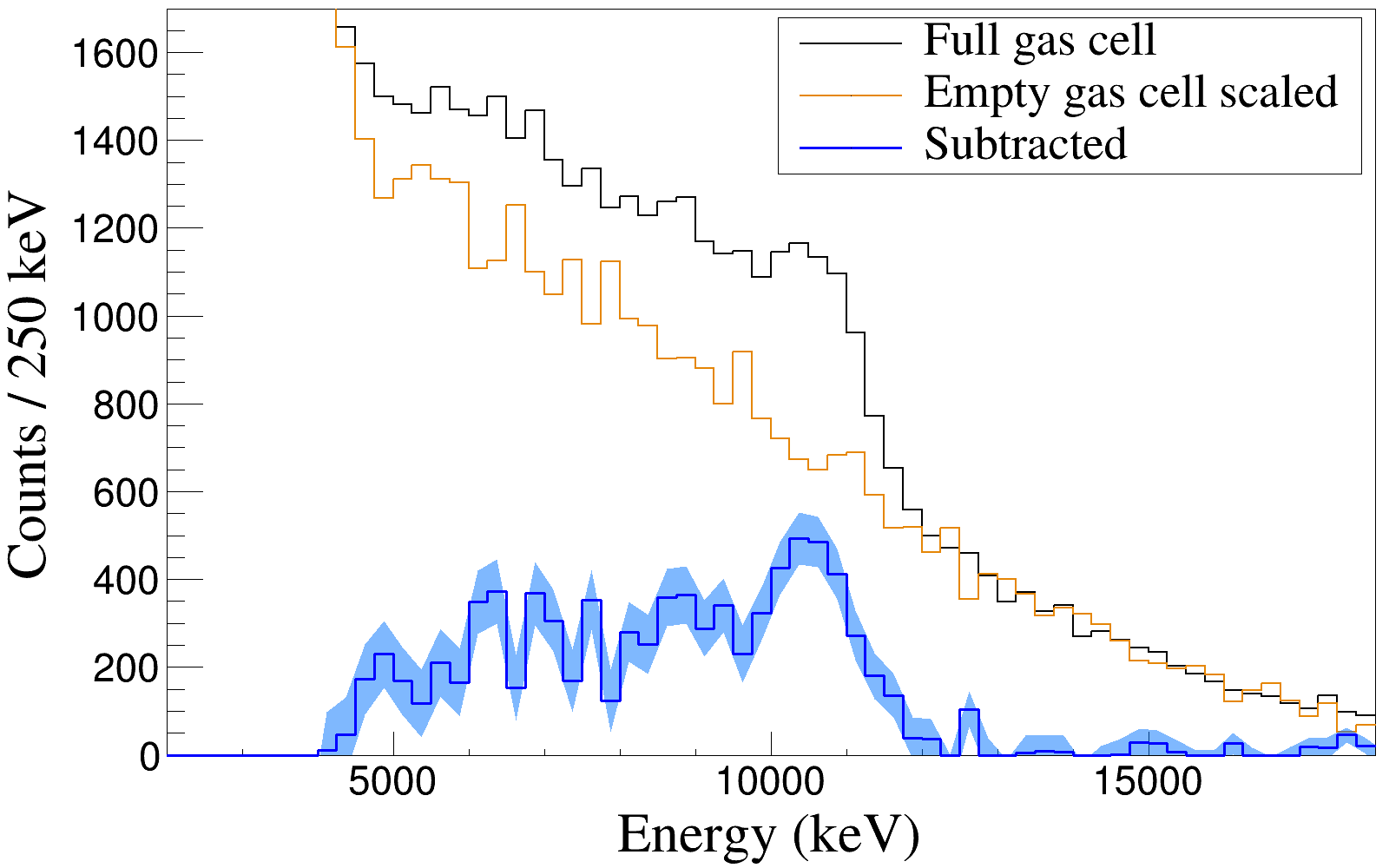}
    \caption{Doppler-corrected TAS spectra showing the background subtraction for the sum peak for the $^{73}\text{As}(p,\gamma)^{74}\text{Se}$ reaction at beam energy of 3.7 MeV/nucleon. The black histogram corresponds to the gas cell filled with hydrogen gas, the orange histogram corresponds to the empty gas cell scaled, and the blue histogram is the fully subtracted sum peak with the band indicating the statistical uncertainty of the background subtraction.}
    \label{fig:bgr_sub}
\end{figure}

The sum peak had background contributions from two main sources: cosmic-ray background and beam-induced background. For the former, the Scintillating Cosmic Ray Eliminating ENsemble (SuNSCREEN), an array of nine plastic scintillator bars arranged in a roof-like configuration above SuN, was utilized as a veto detector \cite{SuNSCREEN}. Events recorded by SuN in coincidence with SuNSCREEN were rejected, reducing our cosmic-ray background by approximately a factor of 3. Any remaining cosmic-ray background contributions were removed by utilizing the pulsed structure of the beam. The beam was delivered in 80-$\mu$s-long pulses every 200 ms, and data were recorded continuously with two time gates: one for the beam-on intervals and another for the background data between pulses, which were normalized to the length of the time gates and subtracted. 
The beam-induced background originated mainly from fusion evaporation reactions of the beam with the molybdenum entrance foil. To account for this background, data were collected both with the cell filled with hydrogen gas and with it empty. After the cosmic-ray background subtraction, the empty-cell data were normalized on the high energies of the spectra (between 13 MeV and 18 MeV), where no contributions are expected from the reaction of interest, and then subtracted from the filled-cell data. Finally, Doppler corrections were applied on a segment-by-segment basis, as described in Refs. \cite{quinn_doppler, palmisano, tsantiri}. The final sum peak, along with the filled-cell and scaled empty-cell data for the beam energy of 3.7 MeV/nucleon, is shown in Fig. \ref{fig:bgr_sub}. 

In addition to Doppler broadening, the sum peak broadens due to energy straggling of the beam as it passes through the cell's entrance foil, resulting in the broadened structure shown in Fig. \ref{fig:bgr_sub}. This straggling introduces a distribution of excitation energies $E_X$, which must be considered when calculating the detection efficiency, $\epsilon$. Furthermore, the SuN detector's efficiency depends on the multiplicity and energy of individual $\gamma$ rays of each cascade \cite{SuN_paper2013}. To address these considerations in the efficiency calculation, the $\gamma$-ray deexcitation of each $E_X$ contributing in the sum peak was simulated using the \textsc{Rainier} code \cite{rainier} and modeled to the detector's response function using \textsc{Geant4} \cite{geant}. Further details on this process can be found in Refs.~\cite{palmisano, tsantiri}. To avoid considering only \textit{s}-wave proton capture, the spin and parity distribution of the simulated entry states in \textsc{Rainier} was obtained from \textsc{Talys} 1.96 \cite{talys}. The simulated TAS, SoS, and multiplicity spectra were fit using a $\chi^2$ minimization code \cite{dombos_chi2} to extract each $E_X$'s contribution to the experimental spectra. The ratio of the experimental yield $Y$ to the detector efficiency $\epsilon$ was calculated as a linear combination of integrals of the simulated sum peaks, weighted by each energy's contribution.

The uncertainty associated with the ratio $Y/\epsilon$ was largely dominated by the statistical uncertainty of the background subtraction. For the measurement with beam energy of 3.7 MeV/nucleon, the statistical uncertainty was $\sim$18\%. Unfortunately, the second measurement, with beam energy of 3.1 MeV/nucleon, suffered from very low statistics, and the associated uncertainty was $\sim$72\%. Smaller contributions to the uncertainty included a $\sim$6\% uncertainty introduced from the normalization method of the empty-cell data as well as a $\sim$13\% uncertainty from the different statistical model inputs used in \textsc{Rainier}. The overall uncertainties for $Y/\epsilon$ were $\sim$24\% and $\sim$74\% for the 3.7 and 3.1 MeV/nucleon beam energies, respectively. The efficiency in detecting $\gamma$ rays from the deexcitation of the $^{74}\text{Se}$ compound nucleus was found to be 48(7)\%. More details can be found in Ref.~\cite{tsantiri_thesis}.

The target areal density was calculated from the average gas cell pressure recorded during each measurement and the gas cell's dimension, with an associated uncertainty of $\sim$5\%. The number of projectiles was calculated based on frequent current measurements on a Faraday cup directly upstream of the SuN detector, accounting for the beam charge state of 23$^+$, which introduced an $\sim$8\% uncertainty to the total incident particles, $N_b$.

To account for the thickness of the hydrogen gas target, an effective center-of-mass energy, $E_\text{eff}$, was calculated as described in Ref. \cite{cauldrons}. This energy corresponds to the point within the target where half of the reaction yield is produced. For nonresonant reactions, a linear decrease in cross section can be assumed across the volume of the target, where the beam loses approximately 250 keV, as calculated using \textsc{Srim} \cite{srim}. The slope of the decreasing cross section along this energy range was obtained from statistical model calculations \cite{nonsmoker}. The thickness of the entrance Mo foil was measured using Rutherford backscattering spectrometry at Hope College and was determined to be 1.93(12) mg/cm$^2$, introducing an uncertainty of $\sim$40 keV/nucleon in the beam energy. Additional uncertainty associated with $E_\text{eff}$ includes the energy straggling of the beam as it traverses through the Mo foil ($\sim$2\%) and the hydrogen gas ($\sim$1\%) and the uncertainty of the beam energy delivered in ReA ($\sim$1\%). The uneven energy straggling distribution from \textsc{Srim} led to asymmetric uncertainty in the effective energy.

\begin{figure}
    \centering
    \includegraphics[width=\linewidth]{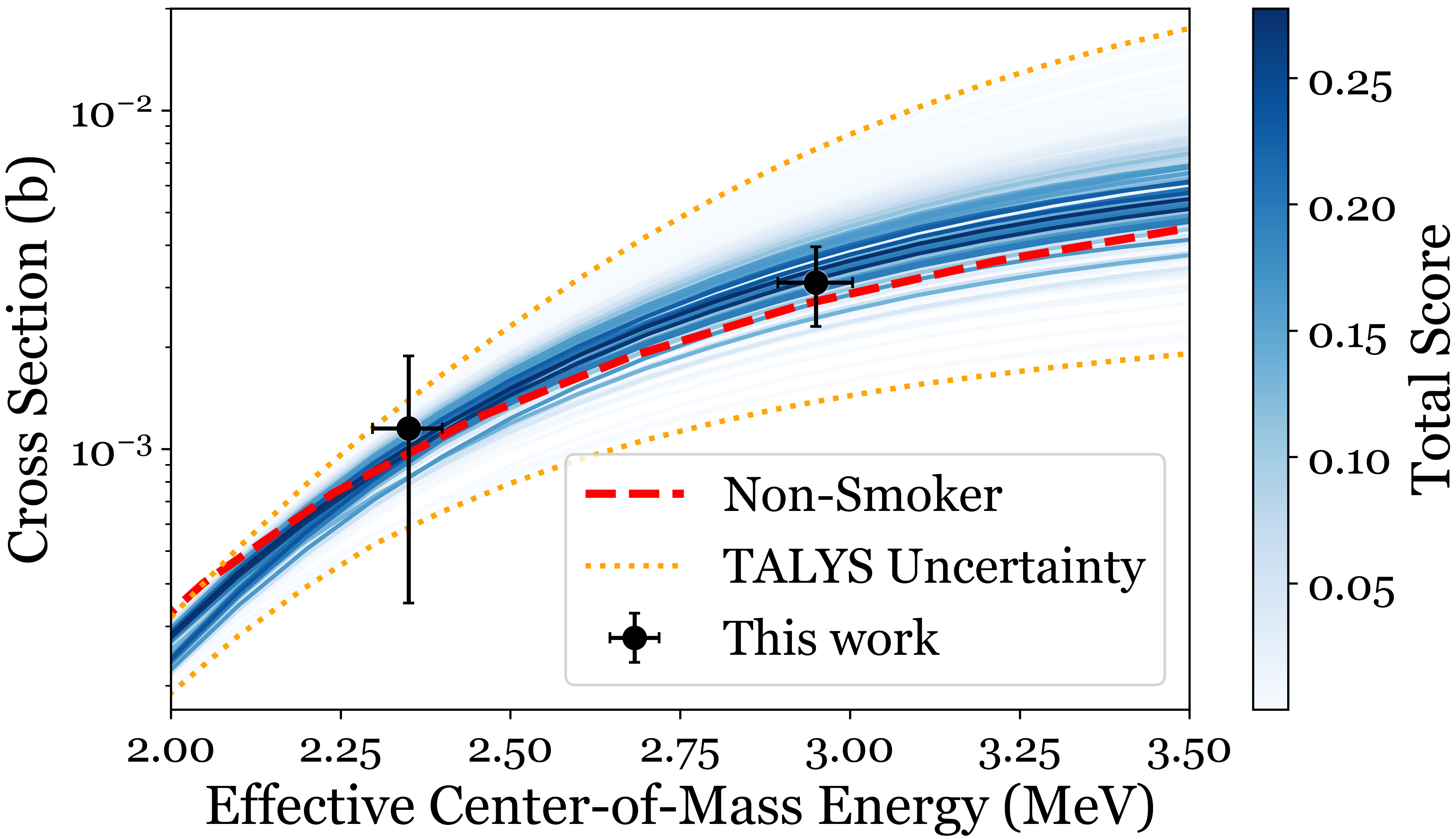}
    \caption{The measured cross section of the $^{73}\text{As}(p,\gamma)^{74}\text{Se}$ reaction (black dots) compared with standard \textsc{Non-Smoker} theoretical calculations \cite{nonsmoker}, represented by a red dashed line, and default \textsc{Talys} 1.96 calculations \cite{talys}, depicted in blue lines. The color coding of the \textsc{Talys} calculations reflects the ability of each model combination to simultaneously describe the experimental spectra and the calculated cross section. More details in text.}    
    \label{fig:cs_plot}
\end{figure}

The cross sections derived from experimental data (Table~\ref{tab:cs_table}) are shown in Fig.~\ref{fig:cs_plot} compared with standard statistical model calculations using the \textsc{Non-Smoker} \cite{nonsmoker} and \textsc{Talys} 1.96 \cite{talys} codes. The color coding of the \textsc{Talys} calculations is discussed in the next paragraphs. The calculated cross sections are in agreement within uncertainty of the \textsc{Non-Smoker} cross section, but the central value is higher than the predictions by $\sim$18\% and $\sim$24\% for the higher and lower energies, respectively. The large statistical uncertainty of the lower energy data point vastly exceeds the spread of the \textsc{Talys} predictions, and therefore, this data point is excluded for the remainder of this analysis. The higher-energy data point provides a strong constraint on the cross section and is used to constrain the statistical properties of the $^{74}\text{Se}$ compound nucleus.

\begin{table*}
    \centering 
    \caption{Measured cross section of the $^{73}\text{As}(p,\gamma)^{74}\text{Se}$ reaction. The first column represents the initial beam energy in the laboratory system, and the second column the center-of-mass effective energy of the reaction. The third column shows the total number of incident particles, the fourth column the total number of reactions that occurred $Y/\epsilon$, and the last shows the measured cross section.}    
    \begin{tabular}{ c @{\hskip 12pt}c@{\hskip 12pt} c@{\hskip 12pt} c@{\hskip 12pt} c@{\hskip 12pt} c}\hline\hline
        \begin{tabular}{@{}c@{}} \rule{0pt}{0.3cm}Initial beam \\ energy (MeV/nucleon)\end{tabular} & \begin{tabular}{@{}c@{}} \rule{0pt}{0.3cm}Effective energy\\ $E_{eff}$ (MeV) \end{tabular} &\begin{tabular}{@{}c@{}} \rule{0pt}{0.3cm}Total incident \\particles $N_b$ \end{tabular}  & \begin{tabular}{@{}c@{}}\rule{0pt}{0.3cm} Total number of \\reactions $Y/\epsilon$ \end{tabular}  &   \begin{tabular}{@{}c@{}}\rule{0pt}{0.3cm}Cross section\\ $\sigma$ (mb) \end{tabular}  \\\hline
        \rule{0pt}{0.5cm}    3.7 & $2.95^{+0.05}_{-0.06}$ & (2.51 $\pm$ 0.21)$\times 10^{10}$ & 6193 $\pm$ 1480 & 3.11 $\pm$ 0.80\\
        \rule{0pt}{0.5cm}    3.1 & 2.35 $\pm$ 0.05 & (1.67 $\pm$ 0.14)$\times 10^{9}$  & $152^{+112}_{-95}$ &  $1.15^{+0.86}_{-0.73}$  \\
        [-2.2ex] &&&&\\\hline\hline
    \end{tabular}
    \label{tab:cs_table}
\end{table*}
The choice of the NLD and $\gamma$SF models in \textsc{Rainier} significantly affects the shape of the simulated spectra, especially the SoS spectrum. Therefore, the ability of the NLD and $\gamma$SF model combinations to describe the simulated compound nucleus is reflected in the $\chi^2$ from the fit of the TAS, SoS, and multiplicity spectra. In Ref.~\cite{tsantiri}, this allowed for the identification of suitable parameter combinations for the analytical descriptions of the NLD and $\gamma$SF. In this work, this methodology was expanded to characterize all possible NLD and $\gamma$SF model combinations available in \textsc{Talys}. 

\begin{figure}
    \centering
    \includegraphics[width=\linewidth]{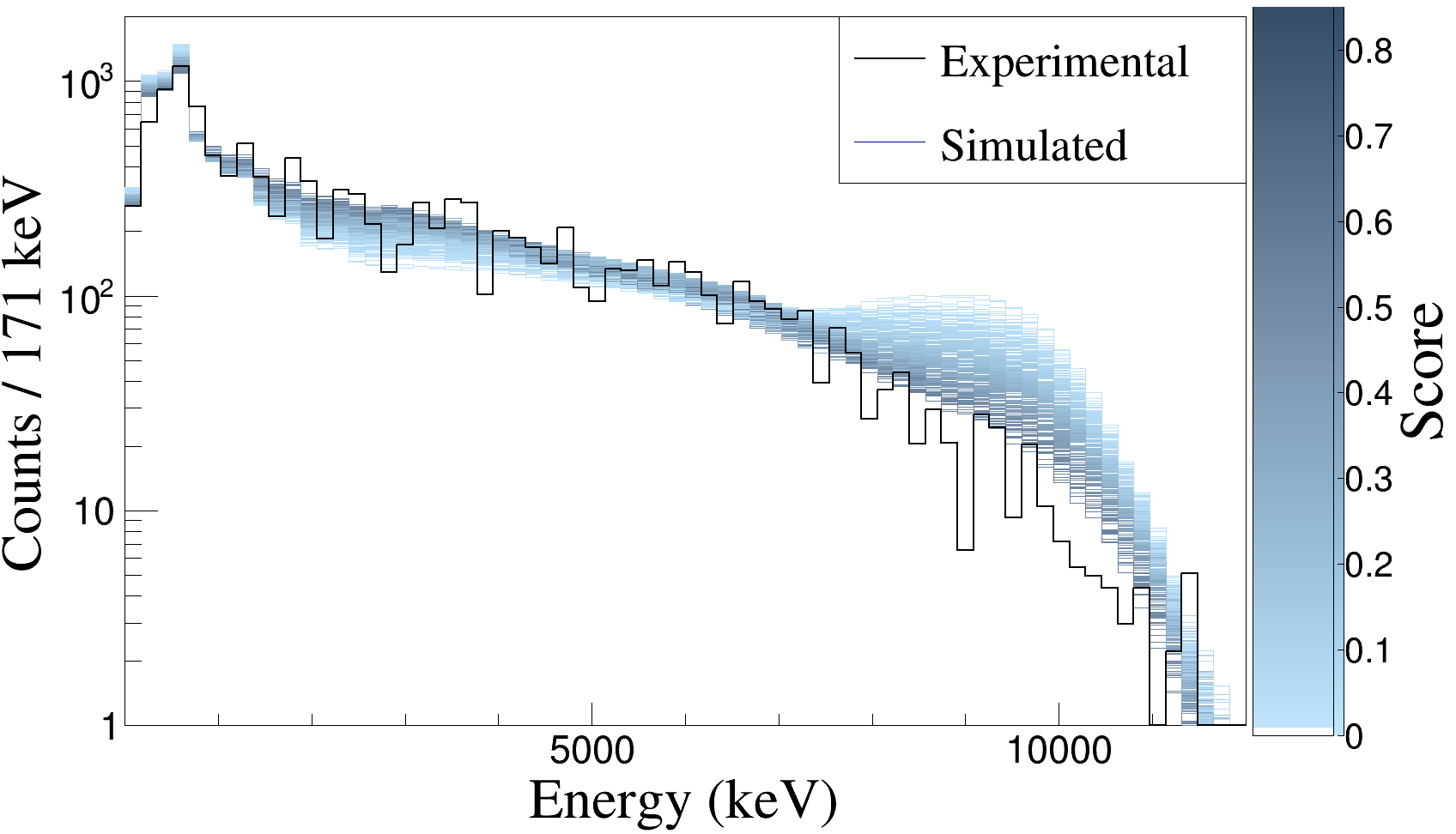}
    \caption{The $\chi^2$ minimization fits of the SoS spectrum for the $^{73}\text{As}(p,\gamma)^{74}\text{Se}$ reaction at a beam energy of 3.7 MeV/nucleon. The black line represents the experimental spectrum, and the various blue lines correspond to the simulated spectra for the combinations of the NLD and $\gamma$SF models from \textsc{Talys} 1.96. The varying shades of blue in each line reflect different scores, where darker tones represent a better description of the experimental spectrum.}    
    \label{fig:chi2_plots}
\end{figure}

Within \textsc{Talys} 1.96 there are six NLD \cite{ctm,bsfg,talys_ld3_ignatyuk, talys_ld3_ignatyuk_2, talys_ld4_goriely, talys_ld5_hilaire, talys_ld6_hilaire}, nine E1  \cite{kopecky_uhl, brink_talys_strength2, axel_talys_strength2, goriely_talys_psf3, goriely_talys_psf4_psf6, goriely_talys_psf5, goriely_talys_psf7, goriely_talys_psf8, plujko_smlo_psf9_talys}, and three M1 \cite{capote_ripl} $\gamma$SF models available, as well as the ability to include a low-energy enhancement (upbend) in the M1 strength function \cite{voinov_upbend, simon_upbend, larsen_upbend_dipole}. The resulting 324 available NLD and $\gamma$SF model combinations were used as input in \textsc{Rainier}, simulated in \textsc{Geant4}, and fit to the experimental TAS, SoS, and multiplicity spectra as described above. The ability of each model combination to reproduce the experimental spectra was reflected in a ``score" calculated based on the $\chi^2$ of each model, as $\exp\left(-\chi^2/2N_\text{bins}\right)$, where a higher score corresponds to a better description of the data. The normalization of the $\chi^2$ to the number of data points (bins) in the respective spectrum, $N_\text{bins}$, was chosen such that each type of spectrum contributed equally to the overall score independently of how many bins it contained, an approach that has been adopted before in the calibration of nuclear models with contributions from multiple data types~\cite{uncertainty_quantified_OP_approach,lovell2017uncertainty,alhassan2022iterative}. As an example, the comparison between the experimental and simulated spectra for the SoS is shown in Fig.~\ref{fig:chi2_plots}. The simulated spectra, color coded based on their scores, correspond to default \textsc{Talys} model combinations. Further investigation of the individual parameters of each model was beyond the scope of this work, as this methodology is sensitive only to the product of NLD and $\gamma$SF and not the individual absolute values.

Aside from their ability to reproduce the experimental spectra, the model combinations were also scored based on their proximity to the obtained cross section, accounting for the uncertainty in cross section and energy. The cross-section uncertainty was assumed to be Gaussian, whereas the energy uncertainty distribution was assumed to follow the energy straggling distribution from \textsc{Srim}. More details on this calculation can be found in Ref.~\cite{tsantiri_thesis}.
Two proton optical model potentials were considered, namely the default \textsc{Talys} parametrization by Koning and Delaroche \cite{koning_delaroche_pomp} as well as the description introduced by Jeukenne, Lejeunne, and Mahaux \cite{Jeu77a,Jeu77b} used by the \textsc{Non-Smoker} code. The overall ability of each of the resulting 648 model combinations to simultaneously describe the experimental TAS, SoS, and multiplicity spectra along with the cross section was calculated from the product of all four scores and is reflected in the color of each \textsc{Talys} calculation in Fig.~\ref{fig:cs_plot}. 

To investigate the impact of this measurement on the production of $^{74}\text{Se}$ in CCSN, Monte Carlo one-zone simulations were performed using the NuGrid code \textsc{ppn} \cite{herwig_2008_nugrid, Pignatari_herwig_2012_nugrid} in a similar manner as in Ref.~\cite{denisseknov_impact_study}. The CCSN model was adapted from Ref.~\cite{ritter_2018_setII} for a 20 $M_\odot$ progenitor, although it should be noted that a similar impact is expected for any CCSN model reaching comparable peak temperatures during the shock wave propagation. In the 20 $M_\odot$ model by Ref.~\cite{ritter_2018_setII}, the maximum production of $^{74}$Se occurs in the inner ONe layer, at mass coordinate $M=2.93\text{ }M_\odot$. Here, the temperature trajectory exhibited a plateau at peak temperature $T=3.08$ GK.

\begin{figure}[t!]
    \centering
    \includegraphics[width=\linewidth]{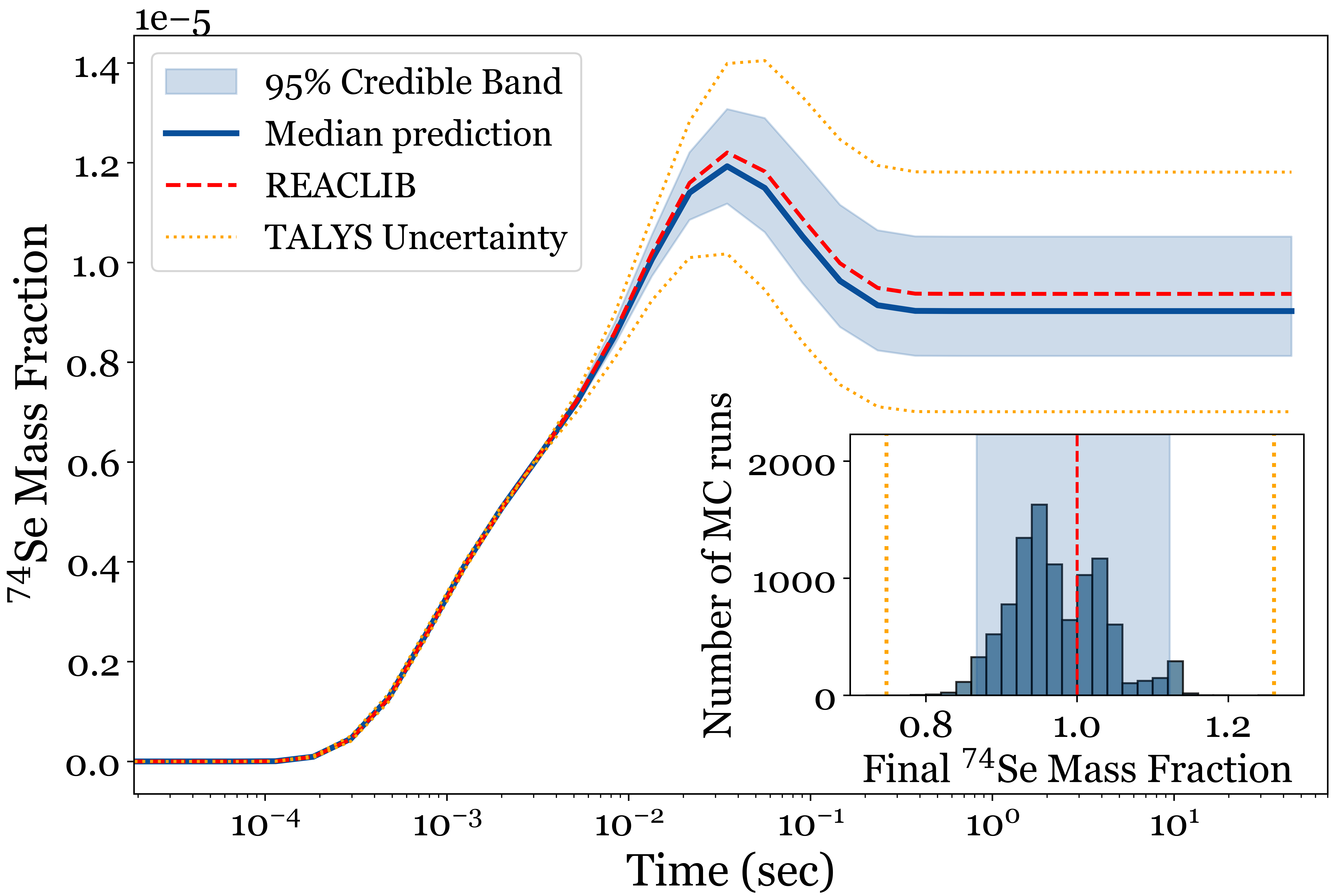}
\caption{Final $^{74}\text{Se}$ mass fraction produced during a 20 $M_\odot$ CCSN as a function of time calculated using NuGrid \textsc{ppn} one-zone simulations \cite{herwig_2008_nugrid, Pignatari_herwig_2012_nugrid} (see text for details). The blue band corresponds to the 95th percentile of the MC sampled $^{73}\text{As}(p,\gamma)^{74}\text{Se}$ reaction rates with the median prediction shown in a darker blue line. The red dashed line corresponds to the default JINA-REACLIB reaction rate \cite{Cyburt_2010_reaclib} and the orange dotted lines to the minimum and maximum reaction rates from \textsc{Talys} 1.96. The inset shows the final $^{74}\text{Se}$ mass fraction distribution obtained in the MC simulations relative to JINA-REACLIB.}
    \label{fig:SNII_74se}
\end{figure}

The \textsc{Talys} cross sections were converted to reaction rates, and for the peak temperature of $T=3.08$ GK, each reaction rate was weighted based on the total score of the corresponding model, an approach akin to Bayesian model averaging~\cite{phillips2021get,hamaker2021precision,neufcourt2020quantified}. From the resulting reaction rate distribution, 10\,000 rates were randomly sampled in a Monte Carlo (MC) method, and in each realization of the \textsc{ppn} code, the rate of the $^{73}\text{As}(p,\gamma)^{74}\text{Se}$ reaction [and its inverse, $^{74}\text{Se}(\gamma,p)^{73}\text{As}$] was adjusted accordingly. The produced $^{74}$Se mass fraction as a function of time is shown in Fig.~\ref{fig:SNII_74se}. The blue band corresponds to the 95th percentile of the MC sampled rates with the median prediction shown in a darker blue line. The red dashed line corresponds to the mass fraction produced using the JINA-REACLIB rate \cite{Cyburt_2010_reaclib} that adopts the \textsc{Non-Smoker} cross section, and the orange dotted lines are the mass fractions obtained with the minimum and maximum rates obtained from \textsc{Talys} 1.96.

As mentioned previously, the experimentally measured $^{73}\text{As}(p,\gamma)^{74}\text{Se}$ cross section used in the current simulations is $\sim$18\% higher than that of \textsc{Non-Smoker}, which is adopted by the JINA-REACLIB database. That being said, the two rates are still consistent within experimental uncertainty, as shown in Fig.~\ref{fig:cs_plot}. Consequently, although a slightly lower mean $^{74}\text{Se}$ production is suggested here, the expected final $^{74}\text{Se}$ abundance from CCSN models remains in good agreement with previous predictions using the JINA-REACLIB rate. Importantly, the presently reported abundance deviation is not enough to suggest that the observed overproduction of $^{74}\text{Se}$ in models of CCSN is driven by uncertainties in the $^{73}\text{As}(p,\gamma)^{74}\text{Se}$ reaction rate. Nevertheless, comparing the full width at half maximum of the distribution of final $^{74}\text{Se}$ mass fractions to the corresponding \textsc{Talys} uncertainty shows that the measurement reduces the uncertainty due to this reaction rate by approximately a factor of 2, providing a more constrained and reliable input for future sensitivity studies. In conclusion, this work strongly constrains the most significant nuclear physics uncertainty related to the synthesis of $^{74}\text{Se}$ and indicates that the overproduction of $^{74}\text{Se}$ points toward the need for a more detailed understanding of the $\gamma$-process conditions such as a different seed distribution, density, temperature, or even a contribution from another process.

In summary, this work presents the first measurement of the $^{73}\text{As}(p,\gamma)^{74}\text{Se}$ reaction cross section using a radioactive $^{73}\text{As}$ beam and only the second radioactive beam experiment of a $\gamma$-process reaction within the Gamow window. The measured cross section was higher than the theoretical prediction from \textsc{Non-Smoker}, but still in agreement within experimental uncertainty. The cross-section data, along with TAS, SoS, and multiplicity spectra, were used to characterize various nuclear level density and $\gamma$-ray strength function models from \textsc{Talys}, allowing for the extraction of an experimentally constrained cross section across the entire Gamow window of the $\gamma$ process. This characterization enabled the determination of an experimentally constrained reaction rate, which was used in Monte Carlo one-zone network simulations of the $\gamma$ process in Type II supernovae. The uncertainty in the final $^{74}\text{Se}$ mass fraction from the $^{73}\text{As}(p,\gamma)^{74}\text{Se}$ reaction is reduced by a factor of two. This study resolves the remaining nuclear physics uncertainties affecting the production of $^{74}$Se and indicates that changes in astrophysical models of the $\gamma$ process are required to explain the solar \textit{p}-nuclei abundances.

\acknowledgments{
The authors would like to thank Aaron Chester and the Facility for Rare Isotope Beams (FRIB) Users Office for their support and assistance during the execution of the experiment and the ReA3 accelerator team for beam delivery. 
Additionally, we would like to thank the NuGrid Collaboration for providing the CCSN models and analysis codes, 
and the International Research Network for Nuclear Astrophysics (IReNA) for enabling this collaboration through the IReNA Visiting Fellowship Program. 
The work was supported by the National Science Foundation under grants
PHY-1913554,  
PHY-2209429, 
PHY-2209138, 
and the U.S. National Nuclear Security Administration through Grant No. DE-NA0004071.
This material is based upon work supported by the U.S. Department of Energy, Office of Science, Office of Nuclear Physics and used resources of FRIB, which is a DOE Office of Science User Facility, operated by Michigan State University, under Award Number DE-SC0023633. 
This work was supported in part by the National Science Foundation under Grant No. OISE-1927130 (IReNA) and the National Nuclear Security Administration under Award No. DE-NA0003180 and the Stewardship Science Academic Alliances program through DOE Awards No DOE-DE-NA0003906.
PD acknowledges support from the Natural Sciences and Engineering Research Council of Canada (NSERC) award SAPPJ-797 2021-00032 ``Nuclear physics of the dynamic origin of the elements".
SL was supported by the Laboratory Directed Research and Development Program at Pacific Northwest National Laboratory operated by Battelle for the U.S. Department of Energy.
}

\bibliography{mybib}

\end{document}